\documentclass[11pt,fleqn]{article}
\usepackage{amsmath,amssymb,lscape}

\newcommand{\R}{{\mathbb R}}
\newcommand{\p}{{\partial}}
\newcommand{\rank}{\mathop{\rm rank}\nolimits}

\newcommand{\vspacebefore}{\raisebox{0ex}[2.5ex][0ex]{\null}}
\newcommand{\myhline}{\\ \hline\vspacebefore}
\newcommand{\Int}{\mathop{\rm Int}\nolimits}

\begin{document}

\allowdisplaybreaks

\begin{flushleft}
\LARGE \bf Computation of Invariants of Lie Algebras\\
 by Means of Moving Frames
\end{flushleft}

\begin{flushleft} \bf
Vyacheslav Boyko~$^\dag$, Jiri Patera~$^\ddag$ and Roman Popovych~$^{\dag\S}$
\end{flushleft}

\noindent $^\dag$~Institute of Mathematics of NAS of Ukraine, 3
Tereshchenkivs'ka Str., Kyiv-4, 01601 Ukraine\\
$\phantom{^\dag}$~E-mail: boyko@imath.kiev.ua, rop@imath.kiev.ua

\noindent
$^\ddag$~Centre de Recherches Math\'ematiques,
Universit\'e de Montr\'eal,\\
$\phantom{^\ddag}$~C.P. 6128 succursale Centre-ville, Montr\'eal (Qu\'ebec), H3C 3J7 Canada\\
$\phantom{^\ddag}$~E-mail: patera@CRM.UMontreal.CA

\noindent $^\S$~Fakult\"at f\"ur Mathematik, Universit\"at Wien, Nordbergstra{\ss}e 15, A-1090 Wien, Austria

\bigskip

\begin{abstract}
\noindent
A new purely algebraic algorithm is presented for computation of
invariants (generalized Casimir operators) of Lie algebras.
It uses the Cartan's method of moving frames and the knowledge of the
group of inner automorphisms of each Lie algebra.
The algorithm is applied, in particular, to computation of invariants of real
low-dimensional Lie algebras. A number of examples are calculated to
illustrate its effectiveness and to make a comparison with the same cases
in the literature. Bases of invariants of
the real solvable Lie algebras up to dimension five,
the real six-dimensional nilpotent Lie algebras and
the real six-dimensional solvable Lie algebras with four-dimensional nilradicals
are newly calculated and listed in tables.
\end{abstract}

\section{Introduction}

Real low-dimensional Lie algebras are finding numerous
applications in many parts of mathematics and physics. Although a
substantive review of these efforts would be desirable, it is
well beyond the scope of the present article. Such
applications provide a general motivation for this work. Result
of a smaller and more specific problem, namely classification of
isomorphism classes of low-dimensional algebras, is the playground
and test bed for the method proposed in this article, although our method
is not constrained to such Lie algebras only (see Example 6 below).

Many authors encountered the need to use a list of isomorphism
classes of the low-dimensional real Lie algebras. In various
degrees of completeness, such lists are available in the literature
\cite{Andrada2004,bianchi1918,Graaf2004,lie1893,MacCallum1999,morozov1958,
mubarakzyanov1963.1,mubarakzyanov1963.2,mubarakzyanov1963.3,
patera&sharp&winternitz1976,patera&winternitz1977,patera&zassenhaus1990b,petrov1966,turkowski1990}
(for review of results on classification of low-dimensional algebras see Table~1 in preprint math-ph/0301029v7).
Unfortunately, it is a laborious and thankless task to unify and correct these lists.
Number of entries in such lists rapidly increases with growing dimension,
even if each parameter-dependent family of non-isomorphic Lie
algebras is counted as a~single entry. Indeed, different choices
of bases of the algebras and ranges of continuous parameters, not
mentioning occasional misprints and errors, make it difficult to
compare such results.
Rigorously speaking, the problem of classification of (solvable) Lie algebras
is \emph{wild} since it includes, as a subproblem, the problem on reduction
of pairs of matrices to a canonical form~\cite{Kirillov}.

The goal of this paper is to introduce an original method for
calculating invariant operators (``generalized Casimir operators'') of the
Lie algebras. In our opinion its main advantage is in that it is
purely algebraic. Unlike the conventional methods, it eliminates
the need to solve systems of differential equations, replacing
them by algebraic equations. Efficient exploitation of the new
method imposes certain constraints on the choice of bases of the
Lie algebras. That then automatically yields simpler expressions
for the invariants. In some cases the simplification is
considerable.

The interest in finding all independent invariants of the real
low-dimensional Lie algebras was recognized a few decades ago
\cite{Abellanas&MartinezAlonso1975,Beltrametti&Blasi1966,
patera&sharp&winternitz1976,Pauri&Prosperi1966,Racah1950,Zassenhaus1977}.
Let us point out that invariants, which are polynomial operators in the
Lie algebra elements, are called here the Casimir operators, while those
which are not necessarily polynomials are called generalized Casimir operators.

At present it looks impossible to construct theory of generalized Casimir
operators in the general case. There are, however, quite a few papers
on properties of such operators, on estimation of their number, on computing
methods and on application of invariants of various classes of Lie algebras,
or even a particular Lie algebra which appears in physical problems.
In particular, functional bases of invariants were calculated for all
three-, \mbox{four-}, five-dimen\-sio\-nal and nilpotent six-dimensional
real Lie algebras in~\cite{patera&sharp&winternitz1976}.
The same problem was considered in~\cite{Ndogmo2000} for the six-dimensional
real Lie algebras with four-dimensional nilradicals.
In~\cite{patera&sharp&winternitz1976a} the subgroups of the Poincar\'e group
together with their invariants were found. The unique (up to independence)
Casimir operator of the unimodular affine group ${\it SA}(4,\mathbb{R})$
which appears, along with the double covering group~$\overline{\it SA}(4,\mathbb{R})$,
as a symmetry group of the spectrum of particles in various gravity-related theories
(metric-affine theory of gravity, particles in curved space-time,
QCD-induced gravity effects on hadrons) is calculated in~\cite{Lemke&Ne'eman&Pecina-Cruz1992}
and then applied to explicit construction of the unitary irreducible representations of
$\overline{\it SA}(4,\mathbb{R})$.

Existence of bases consisting entirely of Casimir operators (polynomial invariants)
is important for the theory of generalized Casimir operators and for their applications.
It was shown that it is the case for the nilpotent and for perfect
Lie algebras~\cite{Abellanas&MartinezAlonso1975}.
A Lie algebra $A$ is perfect if $[A,A]=A$; the derived algebra equals $A$.
(Let us note that the same name is also used for another class of Lie algebras~\cite{Jacobson}.)
Properties of Casimir operators of some perfect Lie algebras and estimations
for their number were investigated recently
in~\mbox{\cite{Campoamor-Stursberg2003c,Campoamor-Stursberg2003e,Ndogmo2004}}.

Invariants of Lie algebras with various additional structural
restrictions are also found in the literature. Namely the solvable Lie algebras
with the nilradicals isomorphic to the Heisenberg algebras~\cite{Rubin&Winternitz1993},
with Abelian nilradicals~\cite{Ndogmo2000a,Ndogmo&Wintenitz1994b},
with nilradicals containing Abelian ideals of codimension~1~\cite{Snobl&Winternitz2005},
solvable triangular algebras~\cite{Tremblay&Winternitz2001},
some solvable rigid Lie algebras~\cite{Campoamor-Stursberg2002a,Campoamor-Stursberg2003b},
solvable Lie algebras with graded nilradical of maximal
nilindex and a Heisenberg subalgebra~\cite{Ancochea2006}.

In~\cite{Perroud1983} the Casimir operators of a number of series of inhomogeneous classical groups
were explicitly constructed. The applied method is based on a particular fiber bundle structure of the generic
orbits generated by the coadjoint representation of a semidirect product.

In this paper, after short review of necessary notions and results,
we formulate a simple algorithm for finding the generalized Casimir operators of Lie algebras.
The algorithm makes use of the Cartan's method of moving frames in the Fels--Olver version
(\cite{Fels&Olver1998,Fels&Olver1999} and reference therein).
It differs from existing methods in that it allows one to avoid integration of systems
of partial differential equations. Then six examples are described in detail.
They are selected to illustrate various aspects of our method.
Finally we present a complete list of corrected and conveniently modified bases of invariants of
the real solvable Lie algebras up to dimension five,
the real six-dimensional nilpotent Lie algebras and
the real six-dimensional solvable Lie algebras with four-dimensional nilradicals.

\section{Preliminaries}

Consider a Lie algebra~$A$ of dimension $\dim A=n<\infty$ over the complex or real field
and the corresponding connected Lie group~$G$.
The results presented in this paper refer to real Lie algebras.

Any (fixed) set of basis elements $e_1,\ldots,e_n$ of~$A$ satisfies the commutation relations
\begin{gather*}
[e_i,e_j]=c_{ij}^k e_k,
\end{gather*}
where $c_{ij}^k$ are components of the tensor of structure constants of~$A$ in the chosen basis.
Hereafter indices $i$, $j$ and~$k$ run from~1 to~$n$ and we use the summation convention for repeated indices.

To introduce the notion of invariants of a Lie algebra, consider the dual space~$A^*$ of the vector space~$A$.
The map ${\rm Ad}^*\colon G\to GL(A^*)$ defined for any $g\in G$ by the relation
$\langle{\rm Ad}^*_g f,a\rangle=\langle f,{\rm Ad}_{g^{-1}}a\rangle$
for all $f\in A^*$ and $a\in A$ is called the {\it coadjoint representation} of the Lie group~$G$.
Here ${\rm Ad}\colon G\to GL(A)$ is the usual adjoint representation of~$G$ in~$A$,
and the image~${\rm Ad}_G$ of~$G$ under~${\rm Ad}$ is the inner automorphism group ${\rm Int}(A)$ of the Lie algebra~$A$.
The image of~$G$ under~${\rm Ad}^*$ is a subgroup of~$GL(A^*)$ and is denoted by~${\rm Ad}^*_G$.

A function $F\in C^\infty(A^*)$ is called an {\it invariant} of~${\rm Ad}^*_G$
if $F({\rm Ad}_g^* f)=F(f)$ for all $g\in G$ and $f\in A^*$.

Our task here is to determine the basis of the functionally independent invariants for
${\rm Ad}^*_G$ and then to transform these invariants to the invariants of the algebra~$A$.
Any other invariant of $A$ is a function of the independent ones.

Let $x=(x_1,\ldots,x_n)$ be the coordinates in~$A^*$ associated with the dual basis to the basis $e_1,\ldots, e_n$.
Any invariant $F(x_1,\ldots,x_n)$ of~${\rm Ad}^*_G$ is a solution of
the linear system of first-order partial differential equations,
see e.g.~\cite{Beltrametti&Blasi1966,Abellanas&MartinezAlonso1975,Pecina-Cruz1994},
\begin{gather}\label{SystemForLieAlgebraInvariants}
X_iF=0,\quad {\rm i.e.}\quad c_{ij}^k x_kF_{x_j}=0,
\end{gather}
where $X_i=c_{ij}^k x_k\partial_{x_j}$ is the infinitesimal generator
of the one-parameter group $\{{\rm Ad}^*_G(\exp\varepsilon e_i)\}$
corresponding to $e_i$. The mapping $e_i\to X_i$ gives a representation of the Lie algebra~$A$.
It is faithful iff the center of $A$ consists of zero only.

It was noted already in \cite{Beltrametti&Blasi1966,Pauri&Prosperi1966}
that the maximal possible number $N_A$ of
functionally independent invariants $F^l(x_1,\ldots,x_n)$, \mbox{$l=1,\ldots,N_A$},
coincides with the number of functionally independent solutions of system~\eqref{SystemForLieAlgebraInvariants}.
It is given by the difference
\begin{gather}\label{N_A}
N_A=\dim A-\rank A.
\end{gather} Here
\[
\rank A=\sup\limits_{(x_1,\ldots,x_n)}\rank\, (c_{ij}^k x_k)_{i,j=1}^n.
\]
The \emph{rank} of the Lie algebra~$A$
 is a bases-independent characteristic of the algebra~$A$.
An interpretation of $N_A$
from the differential form point of view can be found in~\cite{Campoamor-Stursberg2004}.

Given any invariant $F(x_1,\ldots,x_n)$ of~${\rm Ad}^*_G$, one finds
the corresponding invariant of the Lie algebra~$A$ as symmetrization,
$\mathop{\rm Sym}\nolimits F(e_1,\ldots,e_n)$, of $F$.
It is often called a \emph{generalized Casimir operator} of~$A$.
If $F$ is a~polynomial, $\mathop{\rm Sym}\nolimits F(e_1,\ldots,e_n)$ is a usual \emph{Casimir operator}.
More precisely, the symmetrization operator~$\mathop{\rm Sym}\nolimits$ acts only on the monomials
of the forms~$e_{i_1}\cdots e_{i_r}$,
where there are non-commuting elements among~$e_{i_1}, \ldots, e_{i_r}$, and is defined by the formula
\[
\mathop{\rm Sym}\nolimits (e_{i_1}\cdots e_{i_r})=\dfrac1{r!}\sum_{\sigma\in S_r}
e_{i_{\sigma_1}}\cdots e_{i_{\sigma_r}},
\]
where $i_1, \ldots, i_r$ take values from 1 to $n$, $r\in\mathbb{N}$,
the symbol $S_r$ denotes the permutation group of $r$ elements.

The sets of invariants
of ${\rm Ad}^*_G$ and invariants of $A$ are denoted by
$\mathop{\rm Inv}\nolimits({\rm Ad}^*_G)$ and $\mathop{\rm Inv}\nolimits(A)$,
 respectively.
A set of functionally independent invariants
$F^l(x_1,\ldots,x_n)$, \mbox{$l=1,\ldots,N_A$},
forms {\it a~functional basis} ({\it fundamental invariant})
 of $\mathop{\rm Inv}\nolimits({\rm Ad}^*_G)$, i.e.\
any invariant $F(x_1,\ldots,x_n)$ can be uniquely presented as a~function
of~$F^l(x_1,\ldots,x_n)$, \mbox{$l=1,\ldots,N_A$}.
Accordingly the set of $\mathop{\rm Sym}\nolimits F^l(e_1,\ldots,e_n)$, \mbox{$l=1,\ldots,N_A$},
is called a basis of~$\mathop{\rm Inv}\nolimits(A)$.

If the Lie algebra $A$ is decomposable into the direct sum of Lie algebras~$A_1$ and~$A_2$
then the union of bases of~$\mathop{\rm Inv}\nolimits(A_1)$
and~$\mathop{\rm Inv}\nolimits(A_2)$ is a basis of~$\mathop{\rm Inv}\nolimits(A)$.
Therefore, for classification of invariants of Lie algebras from a given class
it is really enough for one to describe only invariants of the indecomposable algebras from this class.

\section{The algorithm}
The standard method
of construction of generalized Casimir operators consists of integration
of the system of linear differential equations~\eqref{SystemForLieAlgebraInvariants}.
It turns out to be rather cumbersome calculation,
once the dimension of Lie algebra is not one of the lowest few.
Alternative methods use matrix representations of Lie algebras, see e.g.\
\cite{Campoamor-Stursberg2005a}.
They are not much easier and are valid for a limited class of representations.

The algebraic method of computation of invariants of Lie algebras presented
in this paper is simpler and generally valid.
It extends to our problem the exploitation of the Cartan's method of moving
frames~\cite{Fels&Olver1998,Fels&Olver1999}.

Let us recall some facts from \cite{Fels&Olver1998,Fels&Olver1999} and adapt them
to the particular case of the coadjoint action of~$G$ on~$A^*$.
Let~$\mathcal{A}={\rm Ad}^*_G\times A^*$ denote the trivial left principal ${\rm Ad}^*_G$-bundle over~$A^*$.
The right regu\-la\-ri\-zation~$\widehat R$ of the coadjoint action of~$G$
on~$A^*$ is the diagonal action of~${\rm Ad}^*_G$ on~$\mathcal{A}={\rm Ad}^*_G\times A^*$. It is provided by the maps
\begin{gather*}
\widehat R_g({\rm Ad}^*_h,f)=({\rm Ad}^*_h\cdot {\rm Ad}^*_{g^{-1}},{\rm Ad}^*_g f),
\qquad g,h\in G, \quad f\in A^*,
\end{gather*}
where the action on the bundle~$\mathcal{A}={\rm Ad}^*_G\times A^*$ is regular and free.
We call $\widehat R_g$ the \emph{lifted coadjoint action} of~$G$. It projects back to the coadjoint
action on~$A^*$ via the ${\rm Ad}^*_G$-equivariant projection~$\pi_{A^*}\colon \mathcal{A}\to A^*$.
Any \emph{lifted invariant} of~${\rm Ad}^*_G$ is a (locally defined) smooth function from~$\mathcal{A}$ to a~manifold,
which is invariant with respect to the lifted coadjoint action of~$G$.
The function $\mathcal{I}\colon\mathcal{A}\to A^*$ given by $\mathcal{I}=\mathcal{I}({\rm Ad}^*_g,f)={\rm Ad}^*_g f$
is the \emph{fundamental lifted invariant} of ${\rm Ad}^*_G$, i.e. $\mathcal{I}$ is a lifted invariant and
any lifted invariant can be locally written as a function of~$\mathcal{I}$.
Using an arbitrary function~$F(f)$ on~$A^*$, we can produce the lifted invariant~$F\circ\mathcal{I}$ of~${\rm Ad}^*_G$
by replacing $f$ with $\mathcal{I}={\rm Ad}^*_g f$ in the expression for~$F$.
Ordinary invariants are particular cases of lifted invariants, where one identifies any invariant formed
as its composition with the standard projection~$\pi_{A^*}$.
Therefore, ordinary invariants are particular functional combinations
of lifted ones that happen to be independent of the group parameters of~${\rm Ad}^*_G$.

In view of the above consideration, the proposed algorithm for construction
of invariants of Lie algebra $A$ can be briefly formulated in the following four steps.

1. {\it Construction of generic matrix $B(\theta)$ of~${\rm Ad}^*_G$.}
It is calculated from the structure constants of the Lie algebra by exponentiation.
$B(\theta)$ is the matrix of an inner automorphism of the Lie algebra~$A$ in the the given basis
$e_1$, \ldots, $e_n$, $\theta=(\theta_1,\ldots,\theta_r)$ are group parameters (coordinates)
of~$\mathop{\rm Int}(A)$, and
\begin{gather*}
r=\dim{\rm Ad}^*_G=\dim\mathop{\rm Int}(A)=n-\dim{\rm Z}(A),
\end{gather*}
${\rm Z}(A)$ is the center of~$A$.
Generally that is a quite straightforward problem if $n=\dim A$ is a~small integer,
and it can be solved by means of using symbolic calculation packages
(we have used Maple~9.0). Computing time may essentially depend on
choose of basis of the Lie algebra $A$.

2. {\it Finite transformations.}
The transformations from ${\rm Ad}^*_G$ can be presented in the coordinate form as
\begin{gather}\label{main}
(\tilde x_1,\ldots,\tilde x_n)=(x_1,\ldots,x_n)\cdot B(\theta_1,\ldots,\theta_r),
\end{gather}
or briefly $\tilde x=x\cdot B(\theta)$. The right-hand member $x\cdot B(\theta)$
of equality~\eqref{main} is the explicit form of the fundamental lifted invariant~$\mathcal{I}$
of ${\rm Ad}^*_G$ in the chosen coordinates~$(\theta,x)$ in ${\rm Ad}^*_G\times A^*$.

3. {\it Elimination of parameters from system~\eqref{main}}.
According to \cite{Fels&Olver1998,Fels&Olver1999}, there are exactly $N_A$~independent
algebraic consequences of~\eqref{main}, which do not contain the parameters $\theta$ ($\theta$-free consequences).
They can be written in the form
\begin{gather*}
F^l(\tilde x_1,\ldots,\tilde x_n)=F^l (x_1,\ldots,x_n),\qquad l=1,\ldots,N_A.
\end{gather*}

4. {\it Symmetrization.} The functions $F^l(x_1,\ldots,x_n)$ which form a basis
of~$\mathop{\rm Inv}\nolimits({\rm Ad}^*_G)$, are symmetrized to
$\mathop{\rm Sym}\nolimits F^l(e_1,\ldots,e_n)$.
It is desired a basis of~$\mathop{\rm Inv}\nolimits(A)$.

\medskip

Let us give some remarks on steps of the algorithm.

In the first step we use second canonical coordinates on $\Int A$ and present the matrix~$B(\theta)$ as
\begin{gather}\label{CalculationOfB}
B(\theta)=\raisebox{0ex}[2ex][1ex]{$\displaystyle\prod_{i=1}^r$}\exp(\theta_i\hat{\rm ad}_{e_{n-r+i}}),
\end{gather}
where $e_1$, \ldots, $e_{n-r}$ are assumed to form a bases of~$Z(A)$;
${\rm ad}_v$ denotes the adjoint representation of $v\!\in\!A$ in~$GL(A)$:
${\rm ad}_v w=[v,w]$ for all $w\!\in\!A$,
and the matrix of ${\rm ad}_v$ in the basis $e_1$, \ldots, $e_n$ is denoted as~$\hat{{\rm ad}}_v$.
In particular, $\hat{{\rm ad}}_{e_i}=(c_{ij}^k)_{j,k=1}^n$.
Sometimes the parameters~$\theta$ are additionally transformed in a light manner (signs, renumbering etc)
for simplification of final presentation of~$B(\theta)$.

Since $B(\theta)$ is a general form of matrices from $\Int A$, we should not adopt it in any way
for the second step.

In fact, the third step of our algorithm involves only preliminaries of the moving frame
method, namely, the procedure of invariant lifting~\cite{Fels&Olver1998,Fels&Olver1999}.
Instead, other closed techniques can be used within the scope of the moving frame method,
which are also based on using an explicit form of finite
transformations~\eqref{main}. One of them is the normalization procedure~\cite{Fels&Olver1998,Fels&Olver1999}.
Following it, we can reformulate the third step of the algorithm.

\medskip

$3'$. {\it Elimination of parameters from lifted invariants}.
We find a nonsingular submatrix
\[
\dfrac{\p(\mathcal{I}_{j_1},\ldots,\mathcal{I}_{j_\rho})}{\p(\theta_{k_1},\ldots,\theta_{k_\rho})}
\qquad (\rho=\rank A)
\]
in the Jacobian matrix~$\p\mathcal{I}/\p\theta$ and solve the equations
$\mathcal{I}_{j_1}=c_1$, \ldots, $\mathcal{I}_{j_\rho}=c_\rho$ with respect to
$\theta_{k_1}$,~\ldots,~$\theta_{k_\rho}$.
Here the constants $c_1$, \ldots, $c_\rho$ are chosen to lie in the range of values of
$\mathcal{I}_{j_1}$, \ldots, $\mathcal{I}_{j_\rho}$.
After substituting the found solutions to the other lifted invariants,
we obtain $N_A=n-\rho$ usual invariants $F^l (x_1,\ldots,x_n)$.

\medskip

\looseness=-1
In conclusion, let us underline that the search of invariants of the Lie algebra $A$,
which has been done by solution of system PDEs \eqref{SystemForLieAlgebraInvariants},
is replaced here by construction of the matrix~$B(\theta)$ of inner automorphisms
and by excluding the parameters~$\theta$ from the algebraic system~\eqref{main} in some way.

\section{Exploitation of the algorithm}

The six examples shown in this Section are selected to give us an opportunity
to make important comments and comparison with analogous results elsewhere.
The Lie algebras are of dimension four, five and six in examples 1--5
and of general finite dimension in the last example.
In some cases the algebras contain continuous parameters,
hence they stand for continuum of non-isomorphic Lie algebras.
For each algebra only the non-zero commutation relations are shown.

Let us point out that simplicity of the form of invariants as well as simplicity of computation often
depend on the choice of bases of Lie algebras. The idea of an appropriate choice of bases
for solvable Lie algebras consists in making evident the chain of
solvable subalgebras $A_i=\langle e_1,\ldots,e_i\rangle$ of ascending dimensions, such that $A_i$ is
an ideal in $A_{i+1}$, $i=1,\ldots,n-1$.
The above basis $e_1$, \ldots, $e_n$ is called $K$-canonical one~\cite{mubarakzyanov1963.1}
which corresponds to the composition series~$K=\{A_i, i=1,\ldots,n\}$.
$K$-canonical bases corresponding to the same composition series~$K$
are connected to each other via linear transformations with triangular matrices.
In particular, we needed to modify bases of solvable Lie algebras in classification
of nilpotent six-dimensional Lie algebras~\cite{morozov1958} and
in classification of six-dimensional Lie algebras with four-dimensional nilradical~\cite{turkowski1990}.

Other criteria of optimality of bases can be used additionally.

\medskip

\noindent
{\bf Example 1.} The Lie algebra in this example is one of the complicate ones among four-dimensional
solvable Lie algebras. On this example we shown all details of our method.

The non-zero commutation relations are the following
\[
[e_1,e_4]=ae_1, \quad [e_2,e_4]=be_2-e_3, \quad [e_3,e_4]=e_2+be_3, \quad a>0, \ \ b\in {\mathbb R}.
\]
It is the Lie algebra $A_{4.6}^{a,b}$
in \cite{mubarakzyanov1963.1,patera&sharp&winternitz1976,
patera&winternitz1977,popovych&boyko&nesterenko&lutfullin2003}. According to \eqref{N_A},
we have $N_A=2$, i.e.\ the algebra~$A_{4.6}^{a,b}$ has two functionally independent invariants.
The matrices of the adjoint representation~$\hat{\rm ad}_{e_i}$ of the basis elements $e_1$, $e_2$, $e_3$ and $e_4$
correspondingly have the form
\[
\left(\begin{array}{cccc} 0&0&0&a\\0&0&0&0\\0&0&0&0\\0&0&0&0\end{array}\right), \quad
\left(\begin{array}{cccc} 0&0&0&0\\0&0&0&b\\0&0&0&-1\\0&0&0&0\end{array}\right), \quad
\left(\begin{array}{cccc} 0&0&0&0\\0&0&0&1\\0&0&0&b\\0&0&0&0\end{array}\right), \quad
\left(\begin{array}{cccc} -a&0&0&0\\0&-b&-1&0\\0&1&-b&0\\0&0&0&0\end{array}\right).
\]

The product of their exponentiations is the matrix $B$ of inner automorphisms of the first step of our algorithm
\begin{gather*}\label{matrixB}
\prod_{i=1}^4\exp(-\theta_i\hat{\rm ad}_{e_i})=B(\theta)=
\left(\begin{array}{cccc} e^{a\theta_4} & 0 & 0 & -a\theta_1 \\
0 & e^{b\theta_4}\cos{\theta_4} & e^{b\theta_4}\sin{\theta_4} & -b\theta_2-\theta_3 \\
0 & -e^{b\theta_4}\sin{\theta_4} & e^{b\theta_4}\cos{\theta_4} & \theta_2-b\theta_3 \\
0 & 0 & 0 & 1\end{array}\right).
\end{gather*}
Substitution $B(\theta)$ into system~\eqref{main} yields the system of linear equations
with coefficients depending explicitly on the parameters $\theta_1$, $\theta_2$, $\theta_3$ and $\theta_4$
\begin{gather*}
\tilde x_1=x_1e^{a\theta_4},\\
\tilde x_2=e^{b\theta_4}(x_2\cos\theta_4 -x_3 \sin\theta_4),\\
\tilde x_3=e^{b\theta_4}(x_2\sin\theta_4 +x_3 \cos\theta_4),\\
\tilde x_4=-ax_1\theta_1 -x_2(b\theta_2+\theta_3)+x_3(\theta_2-b\theta_3)+x_4.
\end{gather*}
Combining the first three equations of the system, one gets
\begin{gather*}
\dfrac{\tilde x_1}{x_1}=e^{a\theta_4},\qquad
\tilde x_2^2+\tilde x_3^2=e^{2b\theta_4}(x_2^2+x_3^2),\qquad
\frac{\tilde x_3}{\tilde x_2}=\tan\left(\arctan\frac{x_3}{x_2}+\theta_4\right).
\end{gather*}
Finally, obvious further combinations lead to the two $\theta$-free relations
\begin{gather*}
\frac{\tilde x_1^b}{(\tilde x_2^2+\tilde x_3^2)^a}=\frac{x_1^b}{(x_2^2+x_3^2)^a},\\[1ex]
(\tilde x_2^2+\tilde x_3^2)\exp\left(-2b\arctan\frac{\tilde x_3}{\tilde x_2}\right)=
(x_2^2+ x_3^2)\exp\left(-2b\arctan\frac{x_3}{x_2}\right).
\end{gather*}
Symmetrization of two expressions does not require any further computation. Consequently, we have
our final results: the two invariants
\begin{gather*}
\frac{e_1^b}{(e_2^2+e_3^2)^a}
\qquad\mbox{and}\qquad
(e_2^2+e_3^2)\exp\left(-2b\arctan\frac{e_3}{e_2}\right)
\end{gather*}
which form a basis of~$\mathop{\rm Inv}\nolimits(A_{4.6}^{ab})$.
It is equivalent to the one constructed in~\cite{patera&sharp&winternitz1976},
but it contains no complex numbers.

\medskip

\noindent
{\bf Example 2.} The solvable Lie algebra~$A_{5.27}$ \cite{mubarakzyanov1963.2,patera&sharp&winternitz1976}
has the following commutation relations~
\[
[e_3,e_4]=e_1,\quad [e_1,e_5]=e_1,\quad [e_2,e_5]=e_1+e_2,\quad [e_3,e_5]=e_2+e_3.
\]
Here we have modified the basis to the $K$-canonical
form~\cite{mubarakzyanov1963.1}, i.e.\ now
$\langle e_1,\ldots, e_i\rangle$ is an ideal in $\langle e_1,\ldots,
e_i, e_{i+1} \rangle$
for any $i=1,2,3,4$. The inner automorphisms of~$A_{5.27}$ are then described by the triangular matrix
\[
B(\theta)=
\left(\begin {array}{ccccc}
e^{\theta_5} & \theta_5 e^{\theta_5} & (\theta_4+\frac 12 \theta_5^2)e^{\theta_5} & \theta_3 & \theta_1+\theta_2 \\
0 & e^{\theta_5} & \theta_5 e^{\theta_5} & 0 & \theta_2+\theta_3 \\
0 & 0 & e^{\theta_5} & 0 & \theta_3\\
0 & 0 & 0 & 1 & 0\\
0 & 0 & 0 & 0 & 1
\end {array}\right).
\]
Combining the first and the second equations of the corresponding system~\eqref{main},
we can exclude the parame\-ter~$\theta_5$ and obtain the relation
\begin{gather*}
\tilde x_1 \exp\left(-\frac{\tilde x_2}{\tilde x_1}\right)=
x_1 \exp\left(-\frac{x_2}{x_1}\right).
\end{gather*}
Since $N_A=1$, there are no other possibilities for construction of $\theta$-free relations.
As a result, we have the basis of~$\mathop{\rm Inv}\nolimits(A_{5.27})$, which consists of the unique element
\[
e_1 \exp\left(-\frac{e_2}{e_1}\right).
\]

\medskip

\noindent
{\bf Example 3.}
The solvable Lie algebra~$A_{5.36}$~\cite{mubarakzyanov1963.2,patera&sharp&winternitz1976}
is defined by the commutation relations
\[
[e_2, e_3] = e_1, \quad [e_1, e_4] = e_1,\quad [e_2, e_4]= e_2, \quad [e_2, e_5] = - e_2,\quad
[e_3,e_5] = e_3.
\]
System \eqref{main} for $A_{5.36}$ has the following form
\begin{gather*}
\tilde x_1=x_1e^{\theta_4},\nonumber\\
\tilde x_2=-x_1\theta_3e^{\theta_4}e^{\theta_5} +x_2e^{\theta_4}e^{\theta_5},\nonumber\\
\tilde x_3=x_1\theta_2e^{-\theta_5} +x_3e^{-\theta_5},\nonumber\\
\tilde x_4=x_1\theta_1 +x_2 \theta_2+x_4,\nonumber\\
\tilde x_5=x_1\theta_2\theta_3-x_2\theta_2+x_3\theta_3+x_5.
\end{gather*}
We multiply the second and third equations and divide the result by the first equation.
Then adding the fifth one, we get \begin{gather*}
\tilde x_5 +\frac{\tilde x_2\tilde x_3}{\tilde x_1}=x_5 +\frac{x_2x_3}{x_1}.
\end{gather*}
Therefore, the right-side member of the latter equality gives an invariant
of coadjoint representation of the Lie group corresponding to $A_{5.36}$.
It is unique up to functional independence since $N_A=1$.
After symmetrization, which is quite non-trivial here in contrast to the
other examples,
we obtain the basis of~$\mathop{\rm Inv}\nolimits(A_{ 5.36})$, formed by the
single invariant
 \[
e_5 +\frac{e_2e_3+e_3e_2}{2e_1}.
\]

\medskip

\noindent
{\bf Example 4.} The six-dimensional Lie algebra
$N_{6.16}^{ab}$~\cite{turkowski1990} is given by the following commutation relations
\begin{gather*}
[e_2,e_5]=e_1,\quad [e_3,e_5]=ae_3+e_4,\quad [e_4,e_5]=-e_3+ae_4,\\ [e_1,e_6]=e_1,
\quad [e_2,e_6]=e_1,\quad [e_3,e_6]=be_3,\quad [e_4,e_6]=be_4, \quad a,b\in {\mathbb R}.
\end{gather*}
For unification with the Mubarakzyanov classification of Lie algebras of dimensions
no greater then~4 and simplicity of our calculations, we have changed numbering of the basis elements
in comparison with \cite{turkowski1990}.

The inner automorphisms of~$N_{6.16}^{ab}$ are defined by the block triangular matrix
\[
B(\theta)=
\left(\begin {array}{cccccc}
e^{\theta_6} & \theta_5 e^{\theta_6} & 0 & 0 & \theta_2 & \theta_1 \\
0 & e^{\theta_6} &0 & 0 & 0 & \theta_2\\
0 & 0 & e^{a\theta_5+b\theta_6}\cos\theta_5 & -e^{a\theta_5+b\theta_6}\sin\theta_5 & a\theta_3-\theta_4 & b\theta_3\\
0 & 0 & e^{a\theta_5+b\theta_6}\sin\theta_5 & e^{a\theta_5+b\theta_6}\cos\theta_5 & \theta_3+a\theta_4 & b\theta_4\\
0 & 0 & 0 & 0 & 1 & 0 \\
0 & 0 & 0 & 0 & 0 & 1
\end {array}\right),
\]
i.e. the corresponding system \eqref{main} has the form
\begin{gather*}
\tilde x_1=e^{\theta_6} x_1,\\
\tilde x_2=\theta_5 e^{\theta_6}x_1+e^{\theta_6}x_2,\\
\tilde x_3=e^{a\theta_5+b\theta_6}(x_3\cos\theta_5 +x_4\sin\theta_5 ),\\
\tilde x_4=e^{a\theta_5+b\theta_6}(-x_3\sin\theta_5 +x_4\cos\theta_5),\\
\tilde x_5=\theta_2 x_1 +(a\theta_3 -\theta_4)x_3 +(\theta_3+a\theta_4)x_4+x_5,\\
\tilde x_6=\theta_1 x_1+\theta_2 x_2 + b\theta_3 x_3 +b\theta_4 x_4+x_6.
\end{gather*}
Obviously only the parameters $\theta_5$ and $\theta_6$, present in the first four equations, can be eliminated.
Namely, the expressions of $\theta_5$ and $\theta_6$ from the first and second equations
are substituted into the third and fourth equations. It leads to two $\theta$-free relations defining the invariants:
\begin{gather*}
\frac{\tilde x_3^2+\tilde x_4^2}{\tilde x_1^{2b}}\exp\left(-2a\frac{\tilde x_2}{\tilde x_1}\right)=
\frac{x_3^2+x_4^2}{x_1^{2b}}\exp\left(-2a\frac{x_2}{x_1}\right),\\[1ex]
\frac{\tilde x_2}{\tilde x_1}+\arctan\frac{\tilde x_4}{\tilde x_3}=
\frac{x_2}{x_1}+\arctan\frac{x_4}{x_3}.
\end{gather*}
The symmetrization procedure is trivial in this case.
As a result, we have the basis of invariants for the Lie algebra~$N_{6.16}^{ab}$
\[
\frac{e_3^2+e_4^2}{e_1^{2b}}\exp\left(-2a\frac{e_2}{e_1}\right),\qquad
\frac{e_2}{e_1}+\arctan\frac{e_4}{e_3}.
\]
It is equivalent to the invariants found in~\cite{Ndogmo2000} but is written in much simpler form.

\medskip

{\samepage
\noindent
{\bf Example 5.} The commutation relations of the solvable Lie algebra of $N_{6.25}^{ab}$~\cite{turkowski1990}
need first to be corrected. Their version in \cite{turkowski1990} contains misprints. After that we get
\begin{gather*}
[e_2,e_5]=ae_2, \quad [e_3,e_5]=e_4, \quad [e_4,e_5]=-e_3, \\
[e_2,e_6]=be_2, \quad [e_3,e_6]=e_3, \quad [e_4,e_6]=e_4, \quad [e_5,e_6]=e_1, \quad a,b \in {\mathbb R},
\quad a^2+b^2\ne 0.
\end{gather*}
As in Example 4, we have suitably renumbered the basis elements.

}

After computing the inner automorphism group of~$N_{6.25}^{ab}$, we get the system~\eqref{main}:
\begin{gather}
\tilde x_1= x_1,\nonumber\\
\tilde x_2=e^{a\theta_4+b\theta_5}x_2,\nonumber\\
\tilde x_3=e^{\theta_5}(x_3\cos\theta_4 +x_4\sin\theta_4 )\nonumber,\\
\tilde x_4=e^{\theta_5}(-x_3\sin\theta_4 +x_4\cos\theta_4),\nonumber\\
\tilde x_5=\theta_5 x_1 +a\theta_1x_2 -\theta_3 x_3 +\theta_2x_4+x_5,\nonumber\\
\tilde x_6=-\theta_4 x_1+b\theta_1 x_2 + \theta_2 x_3 +\theta_3 x_4+x_6.\label{systemEx5}
\end{gather}
Here the parameter~$\theta_i$ correspond to $e_{i+1}$, $i=1,\ldots,5$.
The number~$N_A$ of independent invariants of~$N_{6.25}^{ab}$ equals 2.
It is obvious that $e_1$ generating the center~$Z(N_{6.25}^{ab})$ is one of the invariants.
The second invariant is found by elimination of~$\theta_4$ and $\theta_5$
from the second, third and fourth equations of system \eqref{systemEx5}:
\begin{gather*}
\frac{(\tilde x_3^2+\tilde x_4^2)^b}{\tilde x_2^2}\exp\left(2a\arctan\frac{\tilde x_4}{\tilde x_3}\right)=
\frac{(x_3^2+x_4^2)^b}{x_2^2}\exp\left(2a\arctan\frac{x_4}{x_3}\right).
\end{gather*}
Therefore, we have the following basis of~$\mathop{\rm Inv}\nolimits(N_{6.25}^{ab})$:
\[
e_1,\quad \frac{(e_3^2+e_4^2)^b}{e_2^2}\exp\left(2a\arctan\frac{e_4}{e_3}\right).
\]

The three examples above show that, even for higher dimensional algebras of relative complicated structure,
our method admits hand calculations, provided convenient bases are used.

\medskip

\noindent
{\bf Example 6.} In this example the Lie algebra is of general dimension $n<\infty$.
Consider a class formed by Lie algebras of finite dimensions without an upper bound,
namely by the nilpotent Lie algebras ${\mathfrak n}_{n,1}$, $n=3,4,\ldots$,
with the $(n-1)$-dimensional Abelian ideal $\langle e_1,e_2,\ldots,e_{n-1}\rangle$.
The non-zero commutation relations of~${\mathfrak n}_{n,1}$ have the form~\cite{Snobl&Winternitz2005}
\begin{gather*}
[e_k,e_n]=e_{k-1}, \quad k=2,\ldots,n-1.
\end{gather*}
The inner automorphisms of~${\mathfrak n}_{n,1}$ are described by the triangular matrix
\[
B(\theta)=
\left(\begin {array}{ccccccc}
1 & \theta_1 & \frac{1}{2!}\theta_1^2 & \frac{1}{3!}\theta_1^3 & \cdots & \frac{1}{(n-2)!}\theta_1^{n-2} & \theta_2 \\
0 & 1 & \theta_1 & \frac{1}{2!}\theta_1^2 & \cdots & \frac{1}{(n-3)!}\theta_1^{n-3} & \theta_3 \\
0 & 0 & 1 & \theta_1 & \cdots & \frac{1}{(n-4)!}\theta_1^{n-4} & \theta_4 \\
\cdots & \cdots & \cdots & \cdots & \cdots & \cdots & \cdots\\ 0 & 0 & 0 & 0 & \cdots & \theta_1 & \theta_{n-1}\\
0 & 0 & 0 & 0 & \cdots & 1 & 0 \\
0 & 0 & 0 & 0 & \cdots & 0 & 1 \\
\end {array}\right),
\]
i.e. the complete set of lifted invariants has the form
\[
\mathcal{I}_k=\sum_{j=1}^k \frac{1}{(k-j)!}\theta_1^{k-j} x_j, \quad k=1,\ldots,n-1,\qquad
\mathcal{I}_n=x_n+\sum_{j=1}^{n-2} \theta_{j+1} x_{j}.\label{systemEx6}
\]

It is obvious that the basis element~$e_1$ generating the center of~${\mathfrak n}_{n,1}$
is one of the invariants ($\mathcal{I}_1=x_1$).
Other $(n-3)$ invariants are found by the normalization procedure applied
to the lifted invariants~$\mathcal{I}_2$, \ldots, $\mathcal{I}_{n-1}$.
Namely, we solve the equation $\mathcal{I}_2=0$ with respect to $\theta_1$ and
substitute the obtained expression $\theta_1=-x_2/x_1$ to the other $\mathcal{I}$'s.
To construct polynomial invariants finally, we multiply the derived invariants by powers
of the invariant~$x_1$. Since the symmetrization procedure is trivial for this algebra,
we get the following complete set of generalized Casimir operators which are classical Casimir operators:
\[
e_1, \quad \sum_{j=1}^k \frac{(-1)^{k-j}}{(k-j)!}e_1^{j-2}e_2^{k-j} e_j, \quad k=3,\ldots,n-1.
\]
This set completely coincides with the one determined in Lemma 1 of
\cite{Ndogmo&Wintenitz1994b} and Theorem~4 of~\cite{Snobl&Winternitz2005}.

\section{Concluding remarks}

It is likely that the moving frame method, combined with knowledge
of the groups of inner automorphisms, will allow one to investigate
invariants and other characteristics of special classes of Lie algebras,
such as solvable Lie algebras with given structures of nilradicals (see
Example~6 and~\cite{Boyko&Patera&Popovych2006}).

In the course of testing our method of computing the bases of invariants,
we recalculated invariants of real low-dimensional Lie algebras available
in the literature. A detailed account of this work is to be presented
elsewhere. More precisely, a complete verified list of invariants and
other characteristics, such as the groups of inner automorphisms
of low-dimensional Lie algebras of dimension no greater than six, will be
presented in~\cite{boyko&nesterenko&patera&popovych2005}.

The invariants of Lie algebras of dimension 3, 4, 5, as well as nilpotent
Lie algebras of dimension~6 of \cite{patera&sharp&winternitz1976} are
correct. Note however, that using our new method, all invariants can be
written avoiding introduction of complex numbers (cf. Table~1 and~\cite{patera&sharp&winternitz1976}).
As an illustration, compare Example~1 above with the case of the Lie algebra
$A_{4.6}^{ab}$ of~\cite{patera&sharp&winternitz1976}.
The same claims are true for invariants of the six-dimensional solvable Lie algebras with five-dimensional nilradials,
which are calculated in~\cite{Campoamor-Stursberg2005b}.

Our computation of invariants of the six-dimensional solvable Lie
algebras with four-dimensional nilradials is presented in Table~2.
The same algebras were considered in~\cite{Ndogmo2000}.
Besides correcting several misprints/errors (for example in the entries
$N_{6.3}^a$, $N_{6.12}^{ab}$, $N_{6.21}^{a}$, $N_{6.25}^{ab}$, $N_{6.31}$,
$N_{6.35}^{ab}$ and $N_{6.38}$), we found that, for most of the algebras,
the bases of invariants can be reduced to a simpler form just by choosing
another ($K$-canonical) bases of the algebras.

\medskip

\noindent
{\bf Contents of Tables.}
A symbol $A_{n.k}$ in the first column of Table~1 denotes the indecomposable $n$-dimensional Lie algebra
numbered $k$ in the classifications by Mubarakzyanov~\cite{mubarakzyanov1963.1,mubarakzyanov1963.2} if $n\leqslant5$
or in the classification by Morozov~\cite{morozov1958} if the algebra is nilpotent and six-dimensional.
(This algebra numeration slightly differs from that in~\cite{patera&sharp&winternitz1976}).
Analogously, in the first column the symbol from \cite{turkowski1990} is shown,
denoting six-dimensional solvable Lie algebras with four-dimensional nilradials.
If~a symbol has superscripts, it denotes a series of Lie algebras with parameters indicated in the superscripts.
The parameters $a$, $b$, $c$ and $d$ are real. The non-zero commutation relations are in the second column.
In all cases we have renumbered the basis elements of the algebras in comparison
with~\cite{morozov1958,turkowski1990} in order to have bases in $K$-canonical forms.
Bases of invariants are listed in the third column.
Algebras are collected in correspondence with structure of their nilradicals and centers.
The algebras $N_{6.1}$--$N_{6.19}$ contain the Abelian nilradicals ($\sim 4A_1$) and the centers of dimension zero;
the algebras $N_{6.20}$--$N_{6.27}$ have the Abelian nilradicals ($\sim 4A_1$) and one-dimensional centers;
the nilradical of $N_{6.28}$ is isomorphic to~$A_{4.1}$;
the nilradicals of $N_{6.29}$--$N_{6.40}$ are isomorphic to~$A_{3.1}\oplus A_1$.

\medskip

\noindent
{\bf Acknowledgments.} The work was partially supported by the National Science and Enginee\-ring
Research Council of Canada, by MITACS.
The research of R.\,P. was supported by Austrian Science Fund (FWF), Lise Meitner
project M923-N13. Two of us, V.\,B. and R.\,P., are grateful for the hospitality extended to us at the Centre de
Recherches Math\'ematiques, Universit\'e de Montr\'eal.

\newpage

\begin{center}\footnotesize

{\bf Table 1.} Invariants of the real indecomposable Lie algebras up to dimension 5\\ and
nilpotent Lie algebras of dimension~6

\vspace{.8ex}

\renewcommand{\arraystretch}{1.11}

\begin{tabular}{|@{\,\,}l@{\,\,}|@{\,\,}l@{\,\,}|@{\,\,}l@{\,\,}|}
\hline\vspacebefore
\hfil Algebra &\hfil Nonzero commutation relations & \hfil Invariants 
\\ \hline\vspacebefore
$A_{1}$ &   &  $e_1$
\myhline
$A_{2.1}$ &  $[e_1,e_2]=e_1$ &  none
\myhline
$A_{3.1}$ & $[e_2,e_3]=e_1$ & $e_1$
\myhline
$A_{3.2}$ & $[e_1,e_3]=e_1$, $[e_2,e_3]=e_1+e_2$ &   \raisebox{-1.4ex}[0ex][0ex]{$e_1\exp\Bigl(-\dfrac{e_2}{e_1}\Bigr)$}\\[-.3ex]&&
\myhline
$A_{3.3}$ & $[e_1,e_3]=e_1$, $[e_2,e_3]=e_2$ &  $e_2/e_1$
\myhline
$A_{3.4}^a,$  $|a|\leqslant1,$  &
$[e_1,e_3]=e_1$, $[e_2,e_3]=ae_2$ &
 $e_1^{-a}e_2$ \\
 $a\not=0,1$ &&
\myhline
$A_{3.5}^b,$ $b\geqslant 0$ &
 $[e_1,e_3]=be_1-e_2$, &
 \raisebox{-1.6ex}[0ex][0ex]{$(e_1^2+e_2^2)\exp\Bigl(-2b\arctan\dfrac{e_2}{e_1}\Bigr)$}\\& $[e_2,e_3]=e_1+be_2$ &
\myhline
$\mathrm{sl}(2,{\mathbb R})$ & $[e_1,e_2]=e_1$, $[e_2,e_3]=e_3$, $[e_1,e_3]=-2e_2$  & $e_1e_3+e_3e_1+2e_2^2$
\myhline
$\mathrm{so}(3)$ & $[e_2,e_3]=e_1$, $[e_3,e_1]=e_2$, $[e_1,e_2]=e_3$  &$e_1^2+e_2^2+e_3^2$
\myhline
$A_{4.1}$ & $[e_2,e_4]=e_1$,  $[e_3,e_4]=e_2$  & $e_1$, $e_2^2-2e_1e_3$ 
\myhline
$A_{4.2}^a$, $a\ne 0$ &  $[e_1,e_4]=ae_1$, $[e_2,e_4]=e_2$, $[e_3,e_4]=e_2+e_3$ &
\raisebox{-1.6ex}[0ex][0ex]{$\dfrac{e_2^a}{e_1}$, $e_2\exp\Bigl(-\dfrac{e_3}{e_2}\Bigr)$}
\\&&
\myhline
$A_{4.3}$ &$[e_1,e_4]=e_1$, $[e_3,e_4]=e_2$ & \raisebox{-1.4ex}[0ex][0ex]{$e_2$, $e_1\exp\Bigl(-\dfrac{e_3}{e_2}\Bigr)$} \\[-.3ex] && 
\myhline
$A_{4.4}$ & $[e_1,e_4]=e_1$, $[e_2,e_4]=e_1+e_2$, &
\raisebox{-1.6ex}[0ex][0ex]{$e_1\exp\Bigl(-\dfrac{e_2}{e_1}\Bigr)$, $\dfrac{e_2^2-2e_1e_3}{e_1^2}$}\\
&  $[e_3,e_4]=e_2+e_3$ &
\myhline
$A_{4.5}^{a,b,c}$,& $[e_1,e_4]=ae_1$, $[e_2,e_4]=be_2$, $[e_3,e_4]=ce_3$ & \raisebox{-1.6ex}[0ex][0ex]{$\dfrac{e_3^a}{e_1^c}$, $\dfrac{e_2^a}{e_1^b}$}   \\
 $abc\ne 0$ &&
\myhline
$A_{4.6}^{a,b}$, & $[e_1,e_4]=ae_1$, $[e_2,e_4]=be_2-e_3$, &
\raisebox{-1.6ex}[0ex][0ex]{\parbox{60mm}{$(e_2^2+e_3^2)\exp\left(-2b\arctan\dfrac{e_3}{e_2}\right)$,$\dfrac{(e_2^2+e_3^2)^a}{e_1^b}$}}\\
$a>0$ &  $[e_3,e_4]=e_2+be_3$ &
\myhline
$A_{4.7}$ & $[e_2,e_3]=e_1$,  $[e_1,e_4]=2e_1$, &  \raisebox{-1.6ex}[0ex][0ex]{none} \\
& $[e_2,e_4]=e_2$, $[e_3,e_4]=e_2+e_3$ &
\myhline
$A_{4.8}^a$, $|a|\leqslant 1$ & $[e_2,e_3]=e_1$,  $[e_1,e_4]=(1+a)e_1$, & only for $a=-1$:  \\
& $[e_2,e_4]=e_2$, $[e_3,e_4]=ae_3$ & $e_1$, $e_2e_3+e_3e_2-2e_1e_4$
\myhline
$A_{4.9}^a$, $a\geqslant 0$ & $[e_2,e_3]=e_1$, $[e_1,e_4]=2ae_1$, & only for $a=0$: \\
 & $[e_2,e_4]=ae_2-e_3$, $[e_3,e_4]=e_2+ae_3$ & $e_1$, $2e_1e_4+e_2^2+e_3^2$
\myhline
$A_{4.10}$ & $[e_1,e_3]=e_1$, $[e_2,e_3]=e_2$, $[e_1,e_4]=-e_2$, $[e_2,e_4]=e_1$ & none
\myhline
$A_{5.1}$ &  $[e_3,e_5]=e_1$, $[e_4,e_5]=e_2$ &  $e_1$, $e_2$, $e_2e_3-e_1e_4$
\myhline
$A_{5.2}$ & $[e_2,e_5]=e_1$, $[e_3,e_5]=e_2$, $[e_4,e_5]=e_3$ &  $e_1$, $e_2^2-2e_1e_3$, $e_2^3+ 3e_1^2e_4-3e_1e_2e_3$
\myhline
$A_{5.3}$ & $[e_3,e_4]=e_2$, $[e_3,e_5]=e_1$, $[e_4,e_5]=e_3$ &  $e_1$, $e_2$, $e_3^2+
2e_2e_5 -2e_1e_4$
\myhline
$A_{5.4}$ & $[e_2,e_4]=e_1$, $[e_3,e_5]=e_1$ &  $e_1$
\myhline
$A_{5.5}$ & $[e_3,e_4]=e_1$, $[e_2,e_5]=e_1$, $[e_3,e_5]=e_2$ &  $e_1$
\myhline
$A_{5.6}$ & $[e_3,e_4]=e_1$, $[e_2,e_5]=e_1$, $[e_3,e_5]=e_2$, $[e_4,e_5]=e_3$ &  $e_1$
\myhline
$A_{5.7}^{abc}$,  
 & $[e_1, e_5] = e_1$, $[e_2, e_5] = a e_2$,
$[e_3, e_5] = b e_3$, &  \raisebox{-1.6ex}[0ex][0ex]{$\dfrac{e_1^a}{e_2}$,  $\dfrac{e_1^b}{e_3}$, $\dfrac{e_1^c}{e_4}$}  \\
$abc\not =0$ & $[e_4, e_5] = c e_4$ &
\myhline
$A_{5.8}^a$, & $[e_2, e_5] = e_1$, $[e_3, e_5] = e_3$, $[e_4, e_5]=a e_4$ &
 \raisebox{-1.6ex}[0ex][0ex]{$e_1$, $\dfrac{e_3^a}{e_4}$, $e_3\exp\Bigl(-\dfrac{e_2}{e_1}\Bigr)$}\\
 $0<|a|\leqslant1$ &&
 \myhline
$A_{5.9}^{ab}$, & $[e_1, e_5] = e_1$, $[e_2, e_5] = e_1+e_2$, &
\raisebox{-1.6ex}[0ex][0ex]{$\dfrac{e_1^a}{e_3}$, $\dfrac{e_1^b}{e_4}$, $e_1\exp\Bigl(-\dfrac{e_2}{e_1}\Bigr)$} \\
 $0 \not = b \leqslant a$ & $[e_3, e_5]=a e_3$,  $[e_4, e_5] = b e_4$ &
\myhline
$A_{5.10}$  & $[e_2, e_5] = e_1$, $[e_3, e_5] = e_2$, $[e_4, e_5]= e_4$ &
\raisebox{-1.4ex}[0ex][0ex]{$e_1$, $e_2^2-2e_1e_3$, $e_4\exp\Bigl(-\dfrac{e_2}{e_1}\Bigr)$}\\[-.3ex]&&
\myhline
$A_{5.11}^a$, $a\ne0$ & $[e_1, e_5] = e_1$, $[e_2, e_5] = e_1+e_2$,&
\raisebox{-1.6ex}[0ex][0ex]{$\dfrac{e_1^a}{e_3}$, $e_1\exp\Bigl(-\dfrac{e_2}{e_1}\Bigr)$, $\dfrac{e_2^2-2e_1e_3}{e_1^2}$}\\
& $[e_3, e_5]=e_2+ e_3$,  $[e_4, e_5] = a e_4$ &

\\
\hline
\end{tabular}
\end{center}

\newpage

\begin{center}\footnotesize\renewcommand{\arraystretch}{1.11}%
\begin{tabular}{|@{\,\,}l@{\,\,}|@{\,\,}l@{\,\,}|@{\,\,}l@{\,\,}|}
\hline\vspacebefore
\hfil Algebra &\hfil Nonzero commutation relations & \hfil Invariants 
\\ \hline\vspacebefore
$A_{5.12}$ & $[e_1, e_5] = e_1$, $[e_2, e_5] = e_1+e_2$, &
 \raisebox{-1.6ex}[0ex][0ex]{$e_1\exp\Bigl(-\dfrac{e_2}{e_1}\Bigr)$, $\dfrac{e_2^2\!-\!2e_1e_3}{e_1^2}$,
 $\dfrac{e_2^3\!+\!3e_1^2e_4\!-\!3e_1e_2e_3}{e_1^3}$}\\
 & $[e_3, e_5] = e_2+e_3$, $[e_4, e_5] = e_3+e_4$ &
 \myhline
$A_{5.13}^{abc}$, $|a|\leqslant 1$, & $[e_1, e_5] = e_1$, $[e_2, e_5] = a e_2$,  &
 \raisebox{-1.6ex}[0ex][0ex]{$\dfrac{e_1^a}{e_2}$, $\dfrac{e_3^2+e_4^2}{e_1^{2b}}$,  $e_1^c\exp\Bigl(-\arctan\dfrac{e_4}{e_3}\Bigr)$}\\
 $ac \not =0$ & $[e_3, e_5]= b e_3-c e_4$, $[e_4, e_5] = c e_3+b e_4$ &
 \myhline
$A_{5.14}^{a}$ & $[e_2, e_5] = e_1$, $[e_3, e_5] = a e_3-e_4$,  &
 \raisebox{-1.6ex}[0ex][0ex]{$e_1$, $(e_3^2+e_4^2)\exp\Bigl(-2a\dfrac{e_2}{e_1}\Bigr)$, $\dfrac{e_2}{e_1}-\arctan\dfrac{e_4}{e_3}$}   \\
 &$[e_4, e_5] =  e_3+a e_4$ &
 \myhline
$A_{5.15}^{a}$, $|a| \leqslant 1$ & $[e_1, e_5] = e_1$, $[e_2, e_5] =  e_1+e_2$, &
 \raisebox{-1.6ex}[0ex][0ex]{$\dfrac{e_1^a}{e_3}$, $e_1\exp\Bigl(-\dfrac{e_2}{e_1}\Bigr)$, $e_3\exp\Bigl(-\dfrac{e_4}{e_3}\Bigr)$}  \\
 & $[e_3, e_5]=a e_3$, $ [e_4, e_5] =  e_3+a e_4$&
 \myhline
$A_{5.16}^{ab}$, $b\not =0$ & $[e_1, e_5] = e_1$,
 $[e_2, e_5] = e_1+ e_2$,&
 \raisebox{-1.6ex}[0ex][0ex]{$e_1\exp\Bigl(-\dfrac{e_2}{e_1}\Bigr)$, $\dfrac{e_3^2+e_4^2}{e_1^{2a}}$, $\dfrac{e_2}{e_1}- \arctan\dfrac{e_4}{e_3}$}\\
 &  $[e_3, e_5]= a e_3-b e_4$, $[e_4, e_5] = b e_3+a e_4$  &
 \myhline
$A_{5.17}^{abc}$, $c\not =0$ & $[e_1, e_5] = ae_1-e_2$, $[e_2, e_5] = e_1+ a e_2$, &
 \raisebox{-3.8ex}[0ex][0ex]{\parbox{60mm}{
 $(e_1^2+e_2^2)\exp\Bigl(-2a\arctan\dfrac{e_2}{e_1}\Bigr)$,  $\dfrac{(e_1^2+e_2^2)^b}{(e_3^2+e_4^2)^a}$,\\
 $(e_3^2+e_4^2)\exp\Bigl(-2\dfrac{b}{c}\arctan\dfrac{e_4}{e_3}\Bigr)$
 }}\\
 & $[e_3, e_5]= b e_3-c e_4$, $[e_4, e_5] = c e_3+b e_4$ &\\[1.7ex] &&
\myhline
$A_{5.18}^a$, $a\geqslant 0$ & $[e_1, e_5] = a e_1- e_2$, $[e_2, e_5] = e_1+ a e_2$,&
 \raisebox{-3.8ex}[0ex][0ex]{\parbox{60mm}{
 $(e_1^2+e_2^2)\exp\Bigl(-2a\arctan\dfrac{e_2}{e_1}\Bigr)$, $\dfrac{e_1e_4-e_2e_3}{e_1^2+e_2^2}$,\\
 $(e_1^2+e_2^2)\exp\Bigl(-2a\dfrac{e_1e_3+e_2e_4}{e_1^2+e_2^2}\Bigr)$
 }}\\
 & $[e_3, e_5]= e_1+a e_3- e_4$, $[e_4, e_5] = e_2+e_3+a e_4$ & \\[1.7ex] &&
\myhline
$A_{5.19}^{ab}$, $b \not =0$ & $[e_2, e_3] =  e_1$, $[e_1, e_5] = ae_1$, $[e_2, e_5]= e_2$,
 &
 \raisebox{-1.6ex}[0ex][0ex]{$\dfrac{e_1^b}{e_4^a}$}   \\
 & $[e_3, e_5] = (a-1) e_3$, $[e_4, e_5] = b e_4$ &
\myhline
$A_{5.20}^{a}$ & $[e_2, e_3] =  e_1$, $[e_1, e_5] = ae_2$, $[e_2, e_5]= e_2$,
 &
 \raisebox{-1.6ex}[0ex][0ex]{$e_1\exp\Bigl(-a\dfrac{e_4}{e_1}\Bigr)$}   \\
 & $[e_3, e_5] = (a-1) e_3$, $[e_4, e_5] = e_1+ae_4$ &
\myhline
$A_{5.21}$ & $[e_2, e_3] = e_1$, $[e_1, e_5] = 2 e_1$, $[e_2, e_5]= e_2+ e_3$,
&
 \raisebox{-1.6ex}[0ex][0ex]{$\dfrac{e_4^2}{e_1}$}   \\
 & $[e_3, e_5] =  e_3+ e_4$, $[e_4, e_5] = e_4$  &
\myhline
$A_{5.22}$ & $[e_2, e_3] = e_1$, $[e_2, e_5] = e_3$, $[e_4, e_5]=  e_4$ &
 $e_1$
\myhline
$A_{5.23}^a$, $a\not =0$ & $[e_2, e_3] =  e_1$, $[e_1, e_5] = 2 e_1$, $[e_2, e_5]= e_2+e_3$,
 &  \raisebox{-1.6ex}[0ex][0ex]{$\dfrac{e_1^a}{e_4^2}$}   \\
 & $[e_3, e_5] =  e_3$, $[e_4, e_5] = a e_4$ &
\myhline
$A_{5.24}^\pm$ & $[e_2, e_3] = e_1$, $[e_1, e_5] = 2 e_1$, $[e_2, e_5]= e_2+ e_3$,
 &
 \raisebox{-1.6ex}[0ex][0ex]{$e_1\exp\Bigl(\mp 2\dfrac{e_4}{e_1}\Bigr)$}   \\
 & $[e_3, e_5] = e_3$, $[e_4, e_5] =\pm e_1+2 e_4$ &
\myhline
$A_{5.25}^{ab}$, $b \not = 0$ & $[e_2, e_3] = e_1$, $[e_1, e_5] = 2 a e_1$, $[e_2, e_5]= a e_2+ e_3$,
 &
 \raisebox{-1.6ex}[0ex][0ex]{$\dfrac{e_1^b}{e_4^{2a}}$}   \\
 & $[e_3, e_5] = -e_2 +a e_3$, $[e_4, e_5] = b e_4$ &
\myhline
$A_{5.26}^{\pm a}$ & $[e_2, e_3] = e_1$, $[e_1, e_5] = 2 a e_1$, $[e_2, e_5]
= a e_2+ e_3$, &
 \raisebox{-1.6ex}[0ex][0ex]{$e_1\exp\Bigl(\mp 2a\dfrac{e_4}{e_1}\Bigr)$}   \\
 &  $[e_3, e_5] = -e_2 +a e_3$, $[e_4, e_5] = \pm e_1+ 2 a e_4$ &
\myhline
$A_{5.27}$  & $[e_2, e_3] = e_1$, $[e_1, e_5] = e_1$, $[e_3, e_5]= e_3+ e_4$,
 & \raisebox{-1.6ex}[0ex][0ex]{$e_1 \exp\Bigl(-\dfrac{e_4}{e_1}\Bigr)$}  \\
& $[e_4, e_5] = e_1+e_4$ &
\myhline
$A_{5.28}^a$  &  $[e_2, e_3] =  e_1$, $[e_1, e_5] = ae_1$, $[e_2, e_5]= (a-1) e_2$,
 & \raisebox{-1.6ex}[0ex][0ex]{$\dfrac{e_4^a}{e_1}$} \\
& $[e_3, e_5] =  e_3+e_4$, $[e_4, e_5] =  e_4$ &
\myhline
$A_{5.29}$ &  $[e_2, e_4] =  e_1$, $[e_1, e_5] = e_1$, $[e_2, e_5]=  e_2$, $[e_4, e_5] =  e_3$
& $e_3$
\myhline
$A_{5.30}^a$ & $[e_2, e_4] =  e_1$, $[e_3, e_4] =  e_2$, $[e_1, e_5]= (a+1)e_1$,
  &
 \raisebox{-1.6ex}[0ex][0ex]{$\dfrac{(e_2^2-2e_1e_3)^{a+1}}{e_1^{2a}}$}  \\
 & $[e_2, e_5] =  ae_2$, $[e_3, e_5] = (a-1) e_3$, $[e_4, e_5] = e_4$ &
\myhline
$A_{5.31}$ & $[e_2, e_4] =  e_1$, $[e_3, e_4] = e_2$, $[e_1, e_5]= 3 e_1$,
 &
 \raisebox{-1.6ex}[0ex][0ex]{$\dfrac{(e_2^2-2e_1e_3)^3}{e_1^4}$} \\
 &  $[e_2, e_5] = 2 e_2$, $[e_3,e_5] = e_3$, $[e_4, e_5] = e_3+e_4$ &
\myhline
$A_{5.32}^a$ &  $[e_2, e_4] =  e_1$, $[e_3, e_4] = e_2$, $ [e_1, e_5]=  e_1$,  &
\raisebox{-1.6ex}[0ex][0ex]{$e_1^{2a}\exp\dfrac{e_2^2-2e_1e_3}{e_1^2}$}  \\
&$[e_2, e_5] =  e_2$, $[e_3,e_5] = a e_1 + e_3$ &
\myhline
$A_{5.33}^{ab}$,   &  $[e_1, e_4] =  e_1$, $[e_3, e_4] = a e_3$,
$[e_2, e_5]=  e_2$, & \raisebox{-1.6ex}[0ex][0ex]{$\dfrac{e_1^ae_2^b}{e_3}$}  \\
$a^2+b^2\ne0$&  $[e_3, e_5] =   b e_3$ &
\myhline
$A_{5.34}^{a}$ &  $[e_1, e_4] =  a e_1$, $[e_2, e_4] = e_2$, $[e_3, e_4]=  e_3$,
 & \raisebox{-1.6ex}[0ex][0ex]{$\dfrac{e_2^{a}}{e_1}\exp\dfrac{e_3}{e_2}$}  \\
&$[e_1, e_5] =  e_1$, $[e_3,e_5] =  e_2$ &
\\
\hline
\end{tabular}
\end{center}

\newpage

\begin{center}\footnotesize

\renewcommand{\arraystretch}{1.11}

\begin{tabular}{|@{\,\,}l@{\,\,}|@{\,\,}l@{\,\,}|@{\,\,}l@{\,\,}|}
\hline\vspacebefore
\hfil Algebra &\hfil Nonzero commutation relations & \hfil Invariants 
\\ \hline\vspacebefore
$A_{5.35}^{ab}$,  &  $[e_1, e_4] =  a e_1$, $[e_2, e_4] =  e_2$, $[e_3, e_4]= e_3$,
& \raisebox{-1.6ex}[0ex][0ex]{$\dfrac{e_1^2}{(e_2^2+e_3^2)^a}\exp\Bigl(-2b\arctan \dfrac{e_3}{e_2}\Bigr)$}  \\
$a^2+b^2\ne0$ &$[e_1, e_5] = b e_1$, $[e_2, e_5] =-e_3$, $[e_3, e_5] = e_2$ &
\myhline
$A_{5.36}$ &  $[e_2, e_3] =  e_1$, $[e_1, e_4] = e_1$, $[e_2, e_4]= e_2$,
& \raisebox{-1.6ex}[0ex][0ex]{$ \dfrac{e_2e_3+e_3e_2+2e_1e_5}{e_1}$}  \\
& $[e_2, e_5] = - e_2$,  $[e_3,e_5] = e_3$&
\myhline
$A_{5.37}$ &  $[e_2, e_3] =  e_1$, $[e_1, e_4] = 2 e_1$, $[e_2, e_4]= e_2$,
& \raisebox{-1.8ex}[0ex][0ex]{$\dfrac{e_2^2+e_3^2+2e_1e_5}{e_1}$}  \\
& $[e_3, e_4] = e_3$, $[e_2,e_5] = - e_3$, $[e_3,e_5] = e_2$ &
\myhline
$A_{5.38}$ &  $[e_1, e_4] =  e_1$, $[e_2, e_5] = e_2$, $[e_4, e_5]= e_3$
& $e_3$
\myhline
$A_{5.39}$ &  $[e_1, e_4] =  e_1$, $[e_2, e_4] =  e_2$, $[e_1, e_5]= -e_2$,
& \raisebox{-1.6ex}[0ex][0ex]{$e_3$}  \\
& $[e_2, e_5] = e_1$, $[e_4,e_5] =  e_3$ &
\myhline
$\mathrm{sl}(2,\R)\, +\hspace{-3.5mm}\subset 2A_{1}$ &  $[e_1,e_2]=2e_1$, $[e_1,e_3]=-e_2$, $[e_2,e_3]=2e_3$,
& \raisebox{-1.6ex}[0ex][0ex]{$\big\{e_1e_4^2-e_2e_4e_5-e_3e_5^2\big\}_{\text{symmetrized}}$}  \\
& $[e_1,e_4]=e_5$, $[e_2,e_4]=e_4$, $[e_2,e_5]=-e_5$, $[e_3,e_5]=e_4$ &
\myhline
$A_{6.1}$ & $[e_3,e_6]=e_1$, $[e_4,e_6]=e_2$, $[e_5,e_6]=e_4$ &
 $e_1$, $e_2$,  $e_1e_4-e_2e_3$, $2e_2e_5-e_4^2$
\myhline
$A_{6.2}$ & $[e_2,e_6]=e_1$, $[e_3,e_6]=e_2$, $[e_4,e_6]=e_3$, &
 $e_1$, $2e_1e_3-e_2^2$, $2e_1e_5-2e_2e_4+e_3^2$, \\
 &  $[e_5,e_6]=e_4$ &  $3e_1^2e_4-3e_1e_2e_3+e_2^3$
\myhline
$A_{6.3}$ & $[e_4,e_5]=e_1$, $[e_4,e_6]=e_2$, $[e_5,e_6]=e_3$ &
 $e_1$, $e_2$, $e_3$,  $e_1e_6+e_3e_4 -e_2e_5$
\myhline
$A_{6.4}$ & $[e_3,e_5]=e_1$, $[e_4,e_6]=e_1$, $[e_5,e_6]=e_2$ &
 $e_1$, $e_2$
\myhline
$A_{6.5}^a$, & $[e_3,e_5]=e_1$, $[e_4,e_5]=ae_2$, $[e_3,e_6]=e_2$,&
 \raisebox{-1.6ex}[0ex][0ex]{$e_1$, $e_2$}  \\
 $a\ne 0$ &  $[e_4,e_6]=e_1$  &
\myhline
$A_{6.6}$ & $[e_4,e_5]=e_1$, $[e_3,e_6]=e_1$, $[e_4,e_6]=e_2$, &
 \raisebox{-1.6ex}[0ex][0ex]{$e_1$, $e_2$}  \\
 &  $[e_5,e_6]=e_3$ &
\myhline
$A_{6.7}$ & $[e_4,e_5]=e_1$, $[e_3,e_6]=e_2$, $[e_4,e_6]=e_3$ &
 $e_1$, $e_2$
\myhline
$A_{6.8}$ & $[e_3,e_5]=e_1$, $[e_4,e_6]=e_2$, $[e_5,e_6]=e_3+e_4$ &
 $e_1$, $e_2$
\myhline
$A_{6.9}$ & $[e_4,e_5]=e_1$, $[e_3,e_6]=e_1$, $[e_4,e_6]=e_2$, &
 \raisebox{-1.6ex}[0ex][0ex]{$e_1$, $e_2$}  \\
 & $[e_5,e_6]=e_4$ &
\myhline
$A_{6.10}^a$, & $[e_3,e_5]=e_1$, $[e_4,e_5]=ae_2$, $[e_3,e_6]=e_2$, &
 \raisebox{-1.6ex}[0ex][0ex]{$e_1$, $e_2$}  \\
 $a\ne 0$ &  $[e_4,e_6]=e_1$, $[e_5,e_6]=e_4$ &
\myhline
$A_{6.11}$ & $[e_5,e_6]=e_4$, $[e_4,e_6]=e_3$, $[e_3,e_6]=e_2$,  &
 \raisebox{-1.6ex}[0ex][0ex]{$e_1$, $e_2$}  \\
&  $[e_4,e_5]=e_1$&
\myhline
$A_{6.12}$ & $[e_2,e_5]=e_1$, $[e_3,e_6]=e_1$, $[e_4,e_6]=e_3$ &
 $e_1$, $2e_1e_4-e_3^2$
\myhline
$A_{6.13}$ & $[e_2,e_5]=e_1$, $[e_3,e_6]=e_1$, $[e_4,e_6]=e_3$, &
 \raisebox{-1.6ex}[0ex][0ex]{$e_1$, $2e_1e_4-e_3^2$}  \\
 &  $[e_5,e_6]=e_2$ &
\myhline
$A_{6.14}^a$, & $[e_2,e_5]=ae_2$, $[e_4,e_5]=e_2$, $[e_3,e_6]=e_1$, &
 \raisebox{-1.6ex}[0ex][0ex]{$e_1$, $e_2^2+ae_3^2-2ae_1e_4$} \\
 $a\ne 0$ &  $[e_4,e_6]=e_3$ &
\myhline
$A_{6.15}$ & $[e_2,e_5]=e_1$, $[e_3,e_6]=e_1$, $[e_4,e_6]=e_3$, &
 \raisebox{-1.6ex}[0ex][0ex]{$e_1$, $2e_1e_4-e_3^2$}  \\
 &  $[e_5,e_6]=e_2+e_4$ &
\myhline
$A_{6.16}$ & $[e_3,e_5]=e_1$, $[e_4,e_5]=e_2$, $[e_2,e_6]=e_1$, &
 \raisebox{-1.6ex}[0ex][0ex]{$e_1$, $e_2^3+3e_1^2e_4-3e_1e_2e_3$} \\
 & $[e_3,e_6]=e_2$, $[e_4,e_6]=e_3$ &
\myhline
$A_{6.17}$ & $[e_2,e_5]=e_1$, $[e_3,e_6]=e_1$, $[e_4,e_6]=e_3$,  &
 \raisebox{-1.6ex}[0ex][0ex]{$e_1$, $2e_1e_4-e_3^2$}  \\
 & $[e_5,e_6]=e_4$ &
\myhline
$A_{6.18}^a$, & $[e_2,e_5]=ae_1$, $[e_4,e_5]=e_2$, $[e_3,e_6]=e_1$, &
 \raisebox{-1.6ex}[0ex][0ex]{$e_1$, $e_2^2+ae_3^2-2ae_1e_4$}  \\
$a\ne 0$ &  $[e_4,e_6]=e_3$, $[e_5,e_6]=e_4$  &
\myhline
$A_{6.19}$ & $[e_4,e_5]=e_1$, $[e_2,e_6]=e_1$, $[e_3,e_6]=e_2$, &
 \raisebox{-1.6ex}[0ex][0ex]{$e_1$, $e_2^2-2e_1e_3$}  \\
&   $[e_4,e_6]=e_3$, $[e_5,e_6]=e_4$ &
\myhline
$A_{6.20}$ & $[e_3,e_5]=e_1$, $[e_4,e_5]=e_2$, $[e_2,e_6]=e_1$, &
 \raisebox{-1.6ex}[0ex][0ex]{$e_1$, $e_2^3+3e_1^2e_4-3e_1e_2e_3$}  \\
 & $[e_3,e_6]=e_2$,  $[e_4,e_6]=e_3$, $[e_5,e_6]=e_4$ &
\myhline
$A_{6.21}$ & $[e_3,e_4]=e_1$, $[e_3,e_5]=e_2$, $[e_4,e_5]=e_3$,&
 \raisebox{-1.6ex}[0ex][0ex]{$e_1$, $e_3^2+2e_1e_5-2e_2e_4$}  \\
 &  $[e_2,e_6]=e_1$, $[e_5,e_6]=e_4$  &
\myhline
$A_{6.22}$ & $[e_3,e_4]=e_1$, $[e_3,e_5]=e_2$, $[e_4,e_5]=e_3$, &
 \raisebox{-1.6ex}[0ex][0ex]{$e_1$, $2e_2^3+3e_1e_3^2 +6 e_1^2e_5-6e_1e_2e_4$} \\
 & $[e_2,e_6]=e_1$, $[e_4,e_6]=e_2$, $[e_5,e_6]=e_4$ &
\\
\hline
\end{tabular}
\end{center}

\newpage

\begin{center}
\footnotesize
{\bf Table 2.} Invariants of the real six-dimensional solvable Lie algebras with four-dimensional nilradials

\vspace{1ex}

\renewcommand{\arraystretch}{1.22}
\begin{tabular}{|@{\,\,}l@{\,\,}|@{\,\,}l@{\,\,}|@{\,\,}l@{\,\,}|}
\hline\vspacebefore
\hfil Algebra &\hfil Non-zero commutation relations & \hfil Invariants
\\ \hline\vspacebefore
$N_{6.1}^{abcd}$ & $[e_1,e_5]=ae_1$, $[e_2,e_5]=be_2$, $[e_4,e_5]=e_4$, &
\raisebox{-3.3ex}[0ex][0ex]{$\dfrac{e_3^c e_4^a}{e_1}$, $\dfrac{e_3^d e_4^b}{e_2}$}
\\
&$[e_1,e_6]=ce_1$, $[e_2,e_6]=d e_2$, $[e_3,e_6]=e_3$,& \\
& $ac\ne 0$, $b^2+d^2\ne 0$& \myhline
$N_{6.2}^{abc}$ & $[e_1,e_5]=ae_1$, $[e_2,e_5]=e_2$, $[e_4,e_5]=e_3$, &
\raisebox{-3.3ex}[0ex][0ex]{$\dfrac{e_2^a e_3^{ac-b}}{e_1}$, $e_2 e_3^c\exp\dfrac{e_4}{e_3}$}
\\
& $[e_1,e_6]=be_1$,
$[e_2,e_6]=c e_2$, $[e_3,e_6]=e_3$, & \\
& $[e_4,e_6]=e_4$, $a^2+b^2\ne 0$ & \myhline
$N_{6.3}^{a}$ & $[e_1,e_5]=e_1$, $[e_2,e_5]=e_2$, $[e_4,e_5]=e_3$, &
\raisebox{-3.3ex}[0ex][0ex]{$e_3\exp\Bigl(-\dfrac{e_2}{e_1}\Bigr)$,
$e_1 \exp\Bigl(-\dfrac{e_4}{e_3}-a\dfrac{e_2}{e_1}\Bigr)$}\\
& $[e_1,e_6]=a e_1$, $[e_2,e_6]=e_1+ae_2$, & \\
& $[e_3,e_6]=e_3$, $[e_4,e_6]=e_4$ &
\myhline
$N_{6.4}^{ab}$ & $[e_1,e_5]=e_1$, $[e_2,e_5]=e_2$, $[e_4,e_5]=e_3$, &
\raisebox{-3.3ex}[0ex][0ex]{\parbox{55mm}{$e_3^{2b}\,(e_1^2+e_2^2)^a \exp \Bigl(-2a\dfrac{e_4}{e_3}\Bigr),$\\[0.2ex]
$e_3\exp\Bigl(a\arctan \dfrac{e_2}{e_1}\Bigr)$}}
\\
&$[e_1,e_6]=e_2$, $[e_2,e_6]=-e_1$, $[e_3,e_6]=ae_3$, &\\
&$[e_4,e_6]=be_3+ae_4$, $a\ne 0$&
\myhline
$N_{6.5}^{ab}$ & $[e_1,e_5]=ae_1$, $[e_3,e_5]=e_3$, $[e_4,e_5]=e_3+e_4$, &
\raisebox{-1.6ex}[0ex][0ex]{$\dfrac{e_2^b e_3^a}{e_1}$, $e_3\exp \Bigl(-\dfrac{e_4}{e_3}\Bigr)$}\\
&$[e_1,e_6]=be_1$, $[e_2,e_6]=e_2$, $ab\ne 0$&
\myhline
$N_{6.6}^{ab}$ & $[e_1,e_5]=ae_1$, $[e_2,e_5]=ae_2$, $[e_3,e_5]=e_3$, &
\raisebox{-3.3ex}[0ex][0ex]{$\dfrac{e_3^a}{e_1}\exp \dfrac{e_2}{e_1}$,
$e_3\exp \Bigl(b\dfrac{e_2}{e_1}-\dfrac{e_4}{e_3}\Bigr)$}
\\
&$[e_4,e_5]=e_3+e_4$, $[e_1,e_6]=e_1$,
$[e_2,e_6]=e_1+e_2$, &\\
& $[e_4,e_6]=be_3$, $a^2+b^2\ne 0$&
\myhline
$N_{6.7}^{abc}$ & $[e_1,e_5]=ae_1$, $[e_2,e_5]=ae_2$, $[e_3,e_5]=e_3$, &
\raisebox{-3.3ex}[0ex][0ex]{\parbox{55mm}{$e_3\exp \Bigl(-\dfrac{e_4}{e_3}-c\arctan \dfrac{e_2}{e_1}\Bigr)$,\\[0.2ex]
$(e_1^2+e_2^2)e_3^{-a}\exp \Bigl(2b\arctan \dfrac{e_2}{e_1}\Bigr)$}}
\\
& $[e_4,e_5]=e_3+e_4$, $[e_1,e_6]=be_1+e_2$, & \\
 & $[e_2,e_6]=-e_1+be_2$, $[e_4,e_6]=ce_3$, $a^2+c^2\ne 0$& \myhline
$N_{6.8}$ & $[e_1,e_5]=e_1$, $[e_4,e_5]=e_2$, $[e_2,e_6]=e_2$, &
\raisebox{-1.6ex}[0ex][0ex]{$e_1\exp \Bigl(-\dfrac{e_4}{e_2}\Bigr)$, $e_2\exp \Bigl(-\dfrac{e_3}{e_2}\Bigr)$}\\
&$[e_3,e_6]=e_2+e_3$, $[e_4,e_6]=e_4$ &
\myhline
$N_{6.9}^a$ & $[e_1,e_5]=e_1$, $[e_4,e_5]=e_2$, $[e_2,e_6]=e_2$, &
\raisebox{-1.6ex}[0ex][0ex]{$e_1^{2a}\exp \biggl(\dfrac{e_3^2-2ae_2e_4}{e_2^2}\biggr)$,
$e_2^a\exp \Bigl(-\dfrac{e_3}{e_2}\Bigr)$}\\
& $[e_3,e_6]=ae_2+e_3$, $[e_4,e_6]=e_3+e_4$&
\myhline
$N_{6.10}^{ab}$ & $[e_1,e_5]=ae_1$, $[e_2,e_5]=e_2$, $[e_3,e_5]=e_3$, &
\raisebox{-3.3ex}[0ex][0ex]{$\dfrac{e_2^{a}}{e_1}\exp\dfrac{e_3}{e_2}$,
$e_2^{2b}\exp \biggl(\dfrac{e_3^2-2e_2e_4}{e_2^2}\biggr)$}\\
&$[e_4,e_5]=be_2+e_4$, $[e_1,e_6]=e_1$, $[e_3,e_6]=e_2$, &\\
& $[e_4,e_6]=e_3$&
\myhline
$N_{6.11}^{a}$ & $[e_2,e_5]=e_1$, $[e_3,e_5]=e_3$, $[e_4,e_5]=e_3+e_4$, &
\raisebox{-1.6ex}[0ex][0ex]{$\dfrac{e_4}{e_3}-\dfrac{e_2}{e_1}$, $\dfrac{e_1^a}{e_3}\exp\dfrac{e_2}{e_1}$}\\
&$[e_1,e_6]=e_1$, $[e_2,e_6]=e_2$, $[e_3,e_6]=ae_3$, $[e_4,e_6]=ae_4$&
\myhline
$N_{6.12}^{ab}$ & $[e_1,e_5]=e_1$, $[e_2,e_5]=e_1+e_2$, $[e_3,e_5]=e_3$,\rule{0ex}{2.7ex} &
\\
& $[e_4,e_5]=e_3+e_4$, $[e_1,e_6]=e_3$, $[e_2,e_6]=ae_1-be_3+e_4$,&
\raisebox{-0ex}[0ex][0ex]{\parbox{60mm}{$\dfrac{e_1e_4-e_2e_3}{e_1^2+e_3^2}+b\arctan\dfrac{e_3}{e_1}$,\\
$\dfrac{e_1e_2+e_3e_4}{e_1^2+e_3^2}+a\arctan\dfrac{e_3}{e_1}+\dfrac 12 \ln (e_1^2+e_3^2)$}} \\
& $[e_3,e_6]=-e_1$, $[e_4,e_6]=be_1-e_2+ae_3$\rule[-1.5ex]{0ex}{2.5ex} &
\myhline
$N_{6.13}^{abcd}$ & $[e_1,e_5]=ae_1$, $[e_2,e_5]=be_2$, $[e_3,e_5]=e_4$, $[e_4,e_5]=-e_3$,
 &
\raisebox{-3.3ex}[0ex][0ex]{\parbox{55mm}{$e_1^2(e_3^2+e_4^2)^{-c}\exp\Bigl(2a\arctan\dfrac{e_4}{e_3}\Bigr)$,\\[0.2ex]
$e_2^2(e_3^2+e_4^2)^{-d}\exp\Bigl(2b\arctan\dfrac{e_4}{e_3}\Bigr)$}}\\
&$[e_1,e_6]=ce_1$, $[e_2,e_6]=de_2$, $[e_3,e_6]=e_3$, $[e_4,e_6]=e_4$,&\\
& $a^2+c^2\ne 0$, $b^2+d^2\ne 0$&
\myhline
$N_{6.14}^{abc}$ & $[e_1,e_5]=ae_1$, $[e_3,e_5]=be_3+e_4$, $[e_4,e_5]=-e_3+be_4$, &
\raisebox{-3.3ex}[0ex][0ex]{\parbox{60mm}{$e_1e_2^{-c}\exp\Bigl(a\arctan\dfrac{e_4}{e_3}\Bigr)$,\\
$(e_3^2+e_4^2)\exp\Bigl(2b\arctan\dfrac{e_4}{e_3}\Bigr)$}}
\\
&$[e_1,e_6]=ce_1$, $[e_2,e_6]=e_2$, $ac\ne 0$& \\
&& \myhline
$N_{6.15}^{abcd}$ & $[e_1,e_5]=e_1$, $[e_2,e_5]=e_2$, $[e_3,e_5]=ae_3+be_4$, &
\raisebox{-3.3ex}[2.8ex][0ex]{\parbox{60mm}{
$(e_1^2+e_2^2)\exp\Bigl(\dfrac{2}{b}\arctan\dfrac{e_4}{e_3}+2c\arctan \dfrac{e_2}{e_1}\Bigr)$,\\[0.1ex]
$(e_3^2+e_4^2)\exp\Bigl(\dfrac{2a}{b}\arctan\dfrac{e_4}{e_3}+2d\arctan \dfrac{e_2}{e_1}\Bigr)$}}
\\
&$[e_4,e_5]=-be_3+ae_4$, $[e_1,e_6]=ce_1+e_2$, &
\\
& $[e_2,e_6]=-e_1+ce_2$, $[e_3,e_6]=de_3$, $[e_4,e_6]=de_4$, $b\ne 0$&
\myhline
$N_{6.16}^{ab}$ & $[e_2,e_5]=e_1$, $[e_3,e_5]=ae_3+e_4$, $[e_4,e_5]=-e_3+ae_4$, &
\raisebox{-1.6ex}[0ex][0ex]{$(e_3^2+e_4^2)e_1^{-2b}\exp\Bigl(-2a\dfrac{e_2}{e_1}\Bigr)$,
$\dfrac{e_2}{e_1}+\arctan\dfrac{e_4}{e_3}$}\\
&$[e_1,e_6]=e_1$, $[e_2,e_6]=e_2$, $[e_3,e_6]=be_3$, $[e_4,e_6]=be_4$ & \myhline
$N_{6.17}^{a}$ & $[e_1,e_5]=ae_1$, $[e_2,e_5]=e_1+ae_2$, $[e_3,e_5]=e_4$, &
\raisebox{-1.6ex}[0ex][0ex]{$e_1\exp\Bigl(-a\dfrac{e_2}{e_1}\Bigr)$, $\dfrac{e_2}{e_1}+\arctan\dfrac{e_4}{e_3}$}\\
&$[e_4,e_5]=-e_3$, $[e_3,e_6]=e_3$, $[e_4,e_6]=e_4$&
\\
\hline
\end{tabular}
\end{center}

\newpage

\begin{center}
\footnotesize

\renewcommand{\arraystretch}{1.11}
\begin{tabular}{|@{\,\,}l@{\,\,}|@{\,\,}l@{\,\,}|@{\,\,}l@{\,\,}|}
\hline\vspacebefore
\hfil Algebra &\hfil Nonzero commutation relations & \hfil Invariants \\ \hline\vspacebefore
$N_{6.18}^{abc}$ & $[e_1,e_5]=e_2$, $[e_2,e_5]=-e_1$, $[e_3,e_5]=ae_3+be_4$, &
\raisebox{-3.2ex}[0ex][0ex]{\parbox{55mm}{$\arctan \dfrac{e_4}{e_3}-b\arctan\dfrac{e_2}{e_1}$,\\
$(e_3^2+e_4^2)(e_1^2+e_2^2)^{-c} \exp\Bigl(2a\arctan\dfrac{e_4}{e_3}\Bigr)$}}
\\
& $[e_4,e_5]=-be_3+ae_4$,
$[e_1,e_6]=e_1$, $[e_2,e_6]=e_2$, &
\\
&$[e_3,e_6]=ce_3$, $[e_4,e_6]=ce_4$, $b\ne 0$\rule[-1.1ex]{0ex}{2ex}&
\myhline
$N_{6.19}$ & $[e_1,e_5]=e_2$, $[e_2,e_5]=-e_1$, $[e_3,e_5]=e_1+e_4$, &
\raisebox{-3.3ex}[0ex][0ex]{$\dfrac{e_1e_4-e_2e_3}{e_1^2+e_2^2}$,
$\dfrac{e_1e_3+e_2e_4}{e_1^2+e_2^2} +\arctan\dfrac{e_4}{e_3}$}
\\
&$[e_4,e_5]=e_2-e_3$, $[e_1,e_6]=e_1$, $[e_2,e_6]=e_2$, &
\\
& $[e_3,e_6]=e_3$, $[e_4,e_6]=e_4$&
\myhline
$N_{6.20}^{ab}$ & $[e_2,e_5]=ae_2$, $[e_4,e_5]=e_4$, $[e_2,e_6]=be_2$, &
\raisebox{-1.6ex}[0ex][0ex]{$e_1$, $\dfrac{e_3^b e_4^a}{e_2}$}
\\
& $[e_3,e_6]=e_3$, $[e_5,e_6]=e_1$&
\myhline
$N_{6.21}^{a}$ & $[e_2,e_5]=e_2$, $[e_4,e_5]=e_3$, $[e_2,e_6]=ae_2$, &
\raisebox{-1.6ex}[0ex][0ex]{$e_1$, $\dfrac{e_3^a}{e_2}\exp\dfrac{e_4}{e_3}$}
\\
& $[e_3,e_6]=e_3$, $[e_4,e_6]=e_4$, $[e_5,e_6]=e_1$&
\myhline
$N_{6.22}^{a\varepsilon}$ & $[e_2,e_5]=e_1$, $[e_4,e_5]=e_4$, $[e_3,e_6]=e_3$, &
\raisebox{-1.6ex}[0ex][0ex]{$e_1$, $\dfrac{e_3^a}{e_4}\exp\dfrac{e_2}{e_1}$}
\\
&$[e_4,e_6]=ae_4$, $[e_5,e_6]=\varepsilon e_1$, $\varepsilon =0,1$, $a^2+\varepsilon^2\ne 0$ &
\myhline
$N_{6.23}^{a \varepsilon}$ & $[e_2,e_5]=e_1$, $[e_3,e_5]=e_3$, $[e_4,e_5]=e_4$, $[e_2,e_6]=ae_1$,&
\raisebox{-1.6ex}[0ex][0ex]{$e_1$, $(e_3^2+e_4^2)\exp\Bigl(-2\dfrac{e_2}{e_1}-2a\arctan \dfrac{e_4}{e_3}\Bigr)$}
\\
&$[e_3,e_6]=e_4$, $[e_4,e_6]=-e_3$, $[e_5,e_6]=\varepsilon e_1$, $\varepsilon =0,1$&
\myhline
$N_{6.24}$ & $[e_3,e_5]=e_3$, $[e_4,e_5]=e_3+e_4$, $[e_2,e_6]=e_2$, $[e_5,e_6]=e_1$\rule[-2.1ex]{0ex}{5.3ex}&
$e_1$, $e_3\exp\Bigl(-\dfrac{e_4}{e_3}\Bigr)$ \myhline
$N_{6.25}^{ab}$ & $[e_2,e_5]=ae_2$, $[e_3,e_5]=e_4$, $[e_4,e_5]=-e_3$,
$[e_2,e_6]=be_2$, &
\raisebox{-1.6ex}[0ex][0ex]{$e_1$, $e_2^2(e_3^2+e_4^2)^{-b}\exp\Bigl(-2a\arctan\dfrac{e_4}{e_3}\Bigr)$}
\\
& $[e_3,e_6]=e_3$, $[e_4,e_6]=e_4$, $[e_5,e_6]=e_1$, $a^2+b^2\ne 0$ &
\myhline
$N_{6.26}^{a}$ & $[e_3,e_5]=ae_3+e_4$, $[e_4,e_5]=-e_3+ae_4$, $[e_2,e_6]=e_2$, &
\raisebox{-1.6ex}[0ex][0ex]{$e_1$, $(e_3^2+e_4^2)\exp\Bigl(2a\arctan\dfrac{e_4}{e_3}\Bigr)$}
\\
& $[e_5,e_6]=e_1$ &
\myhline
$N_{6.27}^{\varepsilon}$ & $[e_2,e_5]=e_1$, $[e_3,e_5]=e_4$, $[e_4,e_5]=-e_3$, $[e_3,e_6]=e_3$, &
\raisebox{-1.6ex}[0ex][0ex]{$e_1$, $\dfrac{e_2}{e_1}+\arctan\dfrac{e_4}{e_3}$}
\\
& $[e_4,e_6]=e_4$, $[e_5,e_6]=\varepsilon e_2$, $\varepsilon =0,1$&
\myhline
$N_{6.28}$ & $[e_2,e_4]=e_1$, $[e_3,e_4]=e_2$, $[e_1,e_5]=e_1$, $[e_3,e_5]=-e_3$, & none \\
& $[e_4,e_5]=e_4$,
$[e_2,e_6]=e_2$, $[e_3,e_6]=2e_3$, $[e_4,e_6]=-e_4$ &
\myhline
$N_{6.29}^{ab}$ & $[e_2,e_3]=e_1$, $[e_1,e_5]=e_1$, $[e_2,e_5]=e_2$, $[e_4,e_5]=ae_4$,
 & none \\
&$[e_1,e_6]=e_1$, $[e_3,e_6]=e_3$, $[e_4,e_6]=be_4$, $a^2+b^2\ne 0$ &
\myhline
$N_{6.30}^{a}$ & $[e_2,e_3]=e_1$, $[e_1,e_5]=2e_1$, $[e_2,e_5]=e_2$, $[e_3,e_5]=e_3$, & none \\
&$[e_4,e_5]=ae_4$, $[e_3,e_6]=e_2$, $[e_4,e_6]=e_4$ &
\myhline
$N_{6.31}$ & $[e_2,e_3]=e_1$, $[e_2,e_5]=e_2$, $[e_3,e_5]=-e_3$, $[e_1,e_6]=e_1$, &
\raisebox{-1.6ex}[0ex][0ex]{$e_1\exp\Bigl(-\dfrac{e_4}{e_1}\Bigr)$, $e_5-\dfrac{e_2e_3+e_3e_2}{2e_1}$}
\\
&$[e_3,e_6]=e_3$, $[e_4,e_6]=e_1+e_4$&
\myhline
$N_{6.32}^a$ & $[e_2,e_3]=e_1$, $[e_2,e_5]=e_2$, $[e_3,e_5]=-e_3$, $[e_4,e_5]=e_1$,
& none \\
&$[e_1,e_6]=e_1$, $[e_2,e_6]=ae_2$, $[e_3,e_6]=(1{-}a)e_3$, $[e_4,e_6]=e_4$ &
\myhline
$N_{6.33}$ & $[e_2,e_3]=e_1$, $[e_1,e_5]=e_1$, $[e_2,e_5]=e_2$,
$[e_1,e_6]=e_1$, & none \\
& $[e_3,e_6]=e_3+e_4$, $[e_4,e_6]=e_4$&
\myhline
$N_{6.34}^a$ & $[e_2,e_3]=e_1$, $[e_1,e_5]=e_1$, $[e_2,e_5]=e_2$, $[e_3,e_5]=e_4$,
 & none \\
&$[e_1,e_6]=a e_1$, $[e_2,e_6]=(a{-}1)e_2$, $[e_3,e_6]=e_3$, $[e_4,e_6]=e_4$ &
\myhline
$N_{6.35}^{ab}$ & $[e_2,e_3]=e_1$, $[e_2,e_5]=e_3$, $[e_3,e_5]=-e_2$, $[e_4,e_5]=ae_4$,&
\raisebox{-3.ex}[0ex][0ex]{\parbox{55mm}{$\dfrac{e_1^b}{e_4^2}$, $2e_5-\dfrac{e_2^2+e_3^2}{e_1}$ if $a=0$,\\[0.5ex]
none if $a\not=0$}}
\\
&$[e_1,e_6]=2e_1$, $[e_2,e_6]=e_2$, $[e_3,e_6]=e_3$, $[e_4,e_6]=be_4$, &\\
& $a^2+b^2\ne 0$ &
\myhline
$N_{6.36}$ & $[e_2,e_3]=e_1$, $[e_2,e_5]=e_3$, $[e_3,e_5]=-e_2$, $[e_1,e_6]=2e_1$, &
\raisebox{-1.6ex}[0ex][0ex]{$e_1\exp\Bigl(-2\dfrac{e_4}{e_1}\Bigr)$, $2e_5-\dfrac{e_2^2+e_3^2}{e_1}$}
\\
&$[e_2,e_6]=e_2$, $[e_3,e_6]=e_3$, $[e_4,e_6]=e_1+2e_4$&
\myhline
$N_{6.37}^a$ & $[e_2,e_3]=e_1$, $[e_2,e_5]=e_3$, $[e_3,e_5]=-e_2$, $[e_4,e_5]=e_1$,
& none \\
&$[e_1,e_6]=2e_1$, $[e_2,e_6]=e_2+ae_3$, $[e_3,e_6]=-ae_2+e_3$, &\\
& $[e_4,e_6]=2e_4$ &
\myhline
$N_{6.38}$ & $[e_2,e_3]=e_1$, $[e_1,e_5]=e_1$, $[e_2,e_5]=e_2$, $[e_1,e_6]=e_1$, &
\raisebox{-1.6ex}[0ex][0ex]{$e_4$, $\dfrac{e_2e_3+e_3e_2}{2e_1}-e_5+e_6+e_4\ln e_1$}
\\
& $[e_3,e_6]=e_3$, $[e_5,e_6]=e_4$&
\myhline
$N_{6.39}$ & $[e_2,e_3]=e_1$, $[e_2,e_5]=e_3$, $[e_3,e_5]=-e_2$, $[e_1,e_6]=2e_1$, &
\raisebox{-1.6ex}[0ex][0ex]{$e_4$, $\dfrac{e_2^2+e_3^2}{e_1}-2e_5+e_4\ln e_1$} \\
&$[e_2,e_6]=e_2$, $[e_3,e_6]=e_3$, $[e_5,e_6]=e_4$ &
\myhline
$N_{6.40}$ & $[e_2,e_3]=e_1$, $[e_2,e_5]=e_3$, $[e_3,e_5]=-e_2$, $[e_4,e_6]=e_4$, &
\raisebox{-1.6ex}[0ex][0ex]{$e_1$, $\dfrac{e_2^2+e_3^2}{e_1}-2e_5+2e_1\ln e_4$} \\
& $[e_5,e_6]=e_1$& \\
\hline
\end{tabular}
\end{center}

\end{document}